\documentclass[%
 reprint,
nofootinbib,
 amsmath,amssymb,
 aps,
 article
]{revtex4-1}

\makeatletter
\def\pdfstartlink@attr{}
\makeatother
\usepackage{hyperref}
\usepackage{graphicx}
\usepackage{dcolumn}
\usepackage{bm}

\usepackage{array}
\usepackage{float}
\usepackage{latexsym}
\usepackage{diagbox}
\usepackage{amsfonts}
\usepackage{cool}
\usepackage{dcolumn}
\usepackage{textcomp}
\usepackage{color}

\usepackage{lipsum}
\usepackage{scrextend}
\usepackage{amsmath}
\usepackage{amssymb}
\usepackage{mathrsfs}

\begin{document}

\preprint{APS/123-QED}

\title{Are fast radio bursts generated by cosmic string cusps?}

\author{Renato Costa}
\email{Renato.Santos@uct.ac.za } 
\author{Jake E. B. Gordin}%
\email{jebgordin@gmail.com} 
\author{Amanda Weltman}%
\email{amanda.weltman@uct.ac.za}
 
\affiliation{The Cosmology and Gravity Group, Department of Mathematics and Applied Mathematics, University of Cape Town, Private Bag, Rondebosch, 7700, South Africa
}%



\date{\today}

\begin{abstract}
We revisit the idea that cosmic strings could source fast radio bursts 
by taking into account Lorentz boosts and a weaker assumption about the scaling law for the energy of particle decay.
We show that the distance relation and time scale, for a specific value of the scaling of energy, are still compatible with observations. However, the event rate predicted by the model is too high when compared to the data. 
We additionally show that a more realistic string, with a finite thickness, further compounds the problem by prohibiting cusp formation and point out how a superconducting wiggly string could circumvent this issue. 

\end{abstract}

\pacs{Valid PACS appear here}
\maketitle


\section{Introduction}

Fast radio bursts (FRBs) are millisecond-duration radio emissions of unknown physical origin (first reported in \cite{Lorimer:2007qn}; for a review, see \cite{Petroff:2017pny,Katz:2016dti}). Among the many attempts to explain FRBs, 
one hypothesis is of interest in this paper: cusp decay from cosmic strings as an FRB mechanism. Electromagnetic signatures from cosmic string cusps at high-energy regimes have been considered before \cite{Brandenberger:1986vj,Bhattacharjee:1989vu,MacGibbon:1989kk,MacGibbon:1992ug,Brandenberger:1993hw}. Recently the idea has been applied to radio frequencies and their cosmological implications \cite{Cai:2011bi,Cai:2012zd,Brandenberger:2017uwo}. 

Standard cosmic strings are described by the Nambu-Goto action and decay primarily through gravitational radiation \cite{Brandenberger:1986vj} (for a review see \cite{Vilenkin:278400,Brandenberger:1993by}). It was previously thought that another signature of cosmic strings would come from non-negligible electromagnetic radiation emitted by cusps \cite{Spergel:1986uu} - portions of the string which double back on themselves 
- due to its considerable size, $l_c \sim r^{1/3} R^{2/3}$, where $r$ is the thickness of the string and $R$ is the radius of the string loop. Given that cosmic strings have yet to be observed, and assuming cusp decay produces radiation across all frequencies, FRBs could, in principle, provide a possible observational testing ground for cosmic strings' signatures, as explored in \cite{Brandenberger:2017uwo}.



There were, however, two assumptions made by the authors in \cite{Brandenberger:2017uwo} which, if dispensed with, alter their conclusions. The first assumption is that the length of the cusp region is unaffected by relativity. Specifically, Lorentz contractions in the rest frame of the string, when taken into account, give a cusp length estimation significantly smaller than the one used in \cite{Brandenberger:2017uwo}. In fact, it was shown in \cite{BlancoPillado:1998bv} that the cusp length actually scales as $l_c \sim \sqrt{rR}$. 

The second assumption is that, when a cusp decays, the number density of particle emission scales as $N(E) \propto E^{-3/2}$. This relation comes from a simply QCD multiplicity function \cite{MacGibbon:1989kk} and is shown to hold well in high energy experiments \cite{Brandenberger:2017uwo}. It is not clear, however, if this holds all the way down to radio energies. As such, we also consider a more general scaling law, $N(E) \propto E^{-m}$, to check whether different values of $m$ can compensate for the much smaller size of the cusp length and still favour the model proposed in \cite{Brandenberger:2017uwo}.

There is a third assumption made in \cite{Brandenberger:2017uwo} and indeed in almost all analyses of electromagnetic radiation from cusps \cite{Brandenberger:1986vj,Bhattacharjee:1989vu,MacGibbon:1989kk,MacGibbon:1992ug,Brandenberger:1993hw}; back-reaction when cusps form is nominally omitted. In Nambu-Goto strings this can be ignored, since with a zero thickness approximation there is no back-reaction. However, from simulations we know that strings in a distribution of string loops form small-scale structure (commonly called wiggles) \cite{Martins:1999su}. These wiggles are dynamical fields coupled to the strings, and so affect the formation of cusps. We show explicitly here that cusp formation is unnatural when a cosmic string is wiggly. 

The paper is divided as follows: in section II, we 
present our analysis assuming the different values for the cusp length and scaling for the energy mentioned above and check how the distance relation, time scale and emission rate change. In section III we consider a more realistic string with a finite thickness given by a mass current and check if cusps can form. We conclude in section IV. 


\section{Cusp length with Lorentz Contractions}

We start by considering the relativistic effects on the cusp length and a more general scaling for the energy of decay particles. The number density of photons from the cusp region, in the limit $E \ll E_f$, is given by 
 \begin{equation}
N(E) \sim \frac{1}{\Theta^2 d^2} \frac{\mu l_c}{{E}^{2}_f} \left(\frac{E_f}{E}\right)^{m},
\end{equation} 
where $E_f$ is the fragmentation energy of the particles produced in the decay process, $d$ is the distance from the cusp at which an observer would receive such a number density of photons and $\Theta$ is the beaming angle. We take $E_f$ to be on the order of the symmetry breaking scale associated with the formation of the cosmic string, \textit{i.e.} $E_f \sim \eta$.
$\mu$ is the mass per unit length of the string, and scales as $\mu \sim \eta^2$. The beaming angle is given by \cite{MacGibbon:1992ug}
\begin{equation}
\Theta \sim \frac{1}{\ln(E_f/ \textnormal{GeV})}.
\end{equation}
We now follow the same procedure as \cite{Brandenberger:2017uwo}. The flux of particles is defined by
\begin{eqnarray}
S &=& \int^E_0 dE' \left(N(E') E'\right) \nonumber \\
&\sim& \frac{\mu l_c}{(2-m)\Theta^2 d^2} \left(\frac{E_f}{E}\right)^{m-2}. 
\end{eqnarray} 
Notice that we must have $m < 2$ in order to have a non-divergent flux. 
Assuming, $E_f \sim \eta$  for the fragmentation energy scale, $l_c \sim r^{1/2}R^{1/2}$ for the cusp length including relativistic corrections \cite{BlancoPillado:1998bv}, and $r \sim \eta^{-1}$ the width of the cosmic string, we have 
\begin{equation}
S \sim \frac{(R\eta)^{-3/2}}{(2-m)\Theta^2 }  \left(\frac{E}{\eta}\right)^{2-m}  \left(\frac{R}{d}\right)^{2} \eta^3.
\end{equation} 
It is useful to express $\eta$ in terms of $m_p$ and $t_0$ (the Planck mass and present time), which are fixed scales. In general, we can write 
\begin{eqnarray}\label{fixetaGmu}
\eta &\sim& (t_0m_p)\frac{\eta}{t_0m_p} \sim  10^{60}\frac{1}{t_0}(G\mu)^{1/2}
\nonumber\\
&\sim& 10^{15} \text{GeV},
\end{eqnarray} 
where in the last line we used $G\mu \sim 10^{-7}$,  $t_0 \sim 10^{42}\text{GeV}^{-1}$. This value for $G\mu$ is the upper bound on the tension of cosmic strings given by CMB measurements of the angular power spectrum \cite{Dvorkin:2011aj,Ade:2013xla,Charnock:2016nzm}. One can, of course, choose small values for the string tension as we will do later.

The energy flux observed is of order $S_0 \sim 10^{-48} (\text{GeV})^3$, so we must have
\begin{eqnarray}
S(E,R,d(R)) &=& S_0. 
\end{eqnarray} 
Using the above relation we can estimate the distance of the source from the detector
\begin{eqnarray}\label{dist}
d(R) &\sim& \frac{R^{1/4}}{\sqrt{(2-m) S_0} \Theta}\bigg(\frac{E}{\eta}\bigg)^{\frac{2-m}{2}} \eta^{3/4}.
\end{eqnarray} 
\subsection{Distance criterion}

The distance \eqref{dist} must be bigger than the expected distance of a cosmic string from us. That has to be so, since the expected distance defined by
\begin{eqnarray}\label{fixetaGmu}
n(R,t)R d_R^3 =1,
\end{eqnarray} 
where $n(R,t)$ is the number density of strings per unit of radius, gives the minimum distance at which one can expect to find at least one cosmic string. It is also the expected mean separation distance of a string loop in a string loop distribution \cite{Brandenberger:2017uwo}.  Using the number density 
\begin{equation}\label{numdensity}
n(R,t) = \beta R^{-5/2} t^{-2} t_{eq}^{1/2}, \qquad \gamma G \mu t < R < t_{eq}
\end{equation}
obtained from the VOS scale model \cite{Kibble:1984hp,Vilenkin:1981kz} in a matter dominated universe\footnote{The number density for a radiation dominated universe is different \cite{Brandenberger:2017uwo}, however we will not consider it here since the signal from strings closer to us will dominate.}, where $\beta$ is a constant of order 1, we get 
\begin{eqnarray}
d_R \sim 10 R^{1/2}t_0^{1/2}.
\end{eqnarray} 

The condition $d_R < d(R)$ implies 
\begin{eqnarray}
R^{1/2} &<&\frac{1\times10^{-2}}{t_0(2-m) \Theta^2  S_0}\bigg(\frac{E}{\eta}\bigg)^{2-m}\hspace{0.5mm} \eta^{3/2}.
\end{eqnarray} \medskip

The fast radio burst energy peak is of order $E\sim 10^{-15} \text{GeV}$. 
For such an $E$, the values for $R$ at different $m$ and $\eta$ are shown in Table \ref{table1}. 
According to the scaling model for the distribution of cosmic string loops \cite{Vilenkin:1981kz,Kibble:1984hp} the radius, in a matter dominated Universe, lies in the interval $\gamma G\mu t_0 < R <  t_{eq}$. Strings with a radius smaller than the lower bound decay in less than one Hubble time. From Table \ref{table1} we see that, considering the same scaling for the energy as 
\cite{Brandenberger:2017uwo}, \textit{i.e.} $m = 3/2$, Lorentz contraction implies a smaller radius for the string loop, which lies outside the allowed region\footnote{In \cite{Brandenberger:2017uwo}, one gets $R \sim 10^{51}\text{(GeV)}^{-1}$ by considering $G\mu \sim 10^{-7}$, which would be inside the allowed region.}. The only way to have a consistent string loop size is if the scaling of the energy is given by $m \sim 2$. A scaling below $m=3/2$ makes the situation even worse. We also compute $R$ for $G\mu \sim 10^{-12}$ and for $G\mu \sim 10^{-18}$, which implies $\eta \sim 10^{12}$ and $\eta \sim 10^{9}$ respectively, to check the impact of string tension on the radius of the loop. Changing the string tension will also change the time scale for the emitted radio burst and the event rate as we will see in the following sections.
\\
\noindent
\begin{table}\label{table1}
\begin{tabular}{| c | c | c | c | c |}
	\hline
    \diagbox{$\eta $}{$m$}  & $\sim$ 2 & 3/2 & 1 & 1/2
    \\
    \hline
    $10^{15}$ \text{GeV} & $\frac{10^{59}}{(2-m)^2}$ & $10^{29}$ &  $10^{-1}$ & $10^{-31}$ \\
	\hline
    $10^{12}$ \text{GeV} & $\frac{10^{50}}{(2-m)^2}$ & $10^{23}$ &  $10^{-4}$  &  $10^{-31}$
    \\
	\hline
$10^{9}$ \text{GeV} & $\frac{10^{42}}{(2-m)^2}$ & $10^{18}$ &  $10^{-8}$  &  $10^{-32}$
     \\
	\hline 
\end{tabular}
\caption{\label{table1}Values of $R$ in $(\text{GeV})^{-1}$ for different $\eta$ and $m$. These values have to be compared to $R_{min}\sim \gamma G\mu t_0$ for differing values of $\eta$. For $\eta \sim 10^{15} \text{GeV}$, $R_{min} \sim 10^{36} \text{GeV}^{-1}$; for $\eta \sim 10^{12} \text{GeV}$, $R_{min} \sim 10^{31} \text{GeV}^{-1}$ and for $\eta \sim 10^{9} \text{GeV}$, $R_{min} \sim 10^{25} \text{GeV}^{-1}$.}
\end{table}


\subsection{Time scale}
The power of cusp decay is given in \cite{Brandenberger:1986vj}
\begin{eqnarray}
P_c \sim \frac{\mu l_c}{R}.
\end{eqnarray} \smallskip
The time scale for a burst can be estimated as the time taken to go from one `side' of the cusp to the other, \textit{i.e.} when two strands of the string at the cusp have moved apart by a distance more than the string width $r$. Since power is energy per unit time, and the string width is inverse of the energy, the time scale for the cusp decay is
\begin{eqnarray}
T = \frac{1}{P_c \hspace{0.5mm}r} \sim R^{1/2} r^{1/2}.
\end{eqnarray} \smallskip
This is different to the time scale in \cite{Brandenberger:2017uwo} by a factor of $(R/r)^{1/6}$. Since the radius of the string loop $R$ varies with the scaling of the energy we will also have different values for the time scale. These are shown in Table \ref{table3}. FRBs are observed as millisecond scale pulses at observation with possibly up to microsecond pulses at emission. 
To be consistent with the distance criterion and time scale of the burst, we must tune $m \sim 2$ and choose $\eta$  of order $10^{12}$ GeV or smaller, as one can check from Table \ref{table1} and Table \ref{table3}. Note that consistency with FRBs data could set another upper bound for the string tension. From the numbers analysed here it would be of order $G\mu \sim 10^{-12}$, which implies an energy scale for the formation of cosmic strings of $\eta \sim 10^{12} \text{GeV}$. Of course a more precise estimation for the string tension would require a more detailed study. We continue with our final consistency check, the event rate.

\begin{table}
\begin{tabular}{| c | c | c | c | c |}
	\hline
    \diagbox{$\eta $}{$m$}  & $\sim$ 2 & 3/2 & 1 
    \\
	\hline
    $10^{15}$ \text{GeV} & $\frac{10^{-2}}{(2-m)}$ & $10^{-17}$ & $10^{-32}$ \\
	\hline
    $10^{12}$ \text{GeV} &  $\frac{10^{-5}}{(2-m)}$  & $10^{-18}$  & $10^{-32}$
    \\
	\hline
    $10^{9}$ \text{GeV} &  $\frac{10^{-8}}{(2-m)}$  & $10^{-20}$  & $10^{-32}$
    \\
	\hline
\end{tabular}
\caption{Time scale of a burst, in seconds, for different values of $m$ and $\eta$.}\label{table3}
\end{table}

\subsection{Event rate}

The event rate of FRBs is given by 
\begin{eqnarray}\label{eventrate}
\mathcal{R} \sim \int^{t_0}_{\gamma G\mu t_0} dR\left(d^3(R) \hspace{0.5mm} n(R, t_0) \frac{1}{R} P(R)\right),
\end{eqnarray} 
where $P(R)$ is the probability of having a beamed jet of photons coming from the cusp in the line of sight. Since we expect one cusp per oscillation, we divide the integrand by $R$. Assuming $P(R)=P_0\sim \Theta$ as approximately constant and plugging (\ref{numdensity}) and (\ref{dist}) into (\ref{eventrate}), we get
\begin{eqnarray}
\mathcal{R} \sim \frac{P_0  (\gamma G\mu)^{-7/4}}{(2-m)^{3/2} \Theta^3}\frac{\beta }{ S_0^{3/2} } \bigg(\frac{E}{\eta}\bigg)^{\frac{3(2-m)}{2}} \eta^{9/4} t_0^{-13/4}10^{-3}.
\nonumber\\
\end{eqnarray}	

The values of $\mathcal{R}$, for different values of $\eta$ and $m$, are shown in Table \ref{table}. Those numbers should be compared to the estimated number of FRBs per year\footnote{Although few FRBs have been detected so far, the expected rate is relevant here.}, which is of order $10^6$.
We can see that if we tune $m$ in order to get a  distance relation and time scale compatible to the data, the number of expected FRBs emitted would be much bigger than the estimated rate. On the other hand, if the scaling for the energy is of order $3/2$ or smaller, the distance relation and event rate are much smaller than the consistency checks require.
This analysis explicitly shows that, unfortunately, FRBs cannot be interpreted as a signature of structureless cosmic strings.

\begin{table}
\begin{tabular}{| c | c | c | c | c |}
	\hline
    \diagbox{$\eta $}{$m$}  & $\sim$ 2 & 3/2 & 1 
    \\
	\hline
    $10^{15}$ \text{GeV} & $\frac{10^{12}}{(2-m)^{3/2}}$ & $10^{-11} $ & $10^{-33}$ \\
	\hline
    $10^{12}$ \text{GeV} &  $\frac{10^{14}}{(2-m)^{3/2}}$  & $10^{-7}$  & $10^{-27}$
    \\
	\hline
    $10^{9}$ \text{GeV} &  $\frac{10^{17}}{(2-m)^{3/2}}$  & $10^{-1}$  & $10^{-19}$
    \\
	\hline
\end{tabular}
\caption{Number of expected FRBs per year for different values of $m$ and $\eta$.}\label{table}
\end{table}

\section{Wiggly strings and cusp formation}

In this section we consider a more realistic string where it could have some small-scale structure by adding wiggles to it. Intuitively, the presence of a mass current could make the situation for FRBs even worse, given that cusps move at the speed of light and wiggles would prevent their formation. We explicitly show that it is the case.

\subsection{Equations of motion}

We start from the Polyakov version of the wiggly string action introduced by \cite{Carter:2003fb}:
\begin{equation}\label{wigglypoly}
S_{wP} = -\frac{M^2}{2} \int d^2\sigma \sqrt{-h}h^{ab} G_{AB} \partial_a X^A\partial_b X^B
\end{equation}
where 
\begin{equation}
G_{AB}=
  \begin{bmatrix}
    g_{\mu\nu} & 0  \\
    0 & \psi^2/M^2
  \end{bmatrix}, \qquad
  X^{A}=
  \begin{bmatrix}
    x^{\mu}   \\
    \phi
  \end{bmatrix}
\end{equation}
and $h^{ab}$ is an auxiliary metric, $h = \det(h_{ij})$, $\phi(\tau, \sigma)$ are the wiggles, $M^2$ is the Kibble  mass scale and $\psi$ is a normalization constant. This action essentially is a NG string coupled dynamically to a scalar field, interpretable as neutral charge carriers (or ``mass'' currents) \cite{Spergel:1986uu,Martins:2014kda}. 

The equation of motion for the above action is obtained by varying it with respect to $X^A$ 
\begin{eqnarray}\label{waveeq}
h^{ab} \partial_a \partial_b X^A &=&0.
\end{eqnarray}
Variation with respect to the auxiliary field $h^{ab}$ gives a set of constraints 
\begin{eqnarray}\label{constraints}
T_{ab} = G_{AB} \partial_a X^A \partial_b X^B -\frac{1}{2} h_{ab} h^{ij} G_{AB}\partial_i X^A \partial_j X^B &=&0.\nonumber
\\
\end{eqnarray}

\subsection{Constraints and the cusp}

We fix world-sheet diffeomorphisms and Weyl invariance by choosing the conformal gauge $h_{ij}=\eta_{ij}$\footnote{Here $\eta_{ij} = diag(-1,1)$ and $c=1$.} and the extra redundancy using the static gauge $x^0 = \tau$. With these choices of gauge, the constraints \eqref{constraints} become
\begin{eqnarray}\label{const1}
-1 + \dot{\textbf{x}} \cdot \dot{\textbf{x}} +  \textbf{x}' \cdot \textbf{x}' &=&- \frac{\psi^2}{M^2}(\dot{\phi}^2+ \phi'^2) ,
\\\label{const2}
\dot{\textbf{x}} \cdot \textbf{x}' &=& - \frac{\psi^2}{2M^2} \dot{\phi}\phi', 
\end{eqnarray}
where $x^\mu=(\tau,\textbf{x})$ and $\dot{A} \equiv \partial_\tau A$, $A' \equiv \partial_\sigma A$ for a general field $A$.
The cusp condition
\begin{eqnarray}
\dot{x}^\mu \dot{x}_\mu = 0 \rightarrow \dot{\textbf{x}} \cdot \dot{\textbf{x}} =1,
\end{eqnarray}
while Lorentz contraction sets $\textbf{x}'=0$ \cite{Spergel:1986uu}. Taking into account the cusp conditions, both constraints can only be satisfied at the same time if, and only if
\begin{eqnarray}
\dot{\phi}(\tau_c,\sigma_c) = \phi'(\tau_c,\sigma_c) =0.
\end{eqnarray}
This condition was also derived in \cite{Spergel:1986uu}, but using a completely different approach. 

The general solution of the equation of motion for $\phi$ have the form
\begin{eqnarray}
\phi(\tau,\sigma) =  f(\sigma - \tau ) +  g( \sigma+\tau ).
\end{eqnarray}
At the cusp, the constraints require 
\begin{eqnarray}
\partial_{-}f(\sigma_{-}) +  \partial_{+}g(\sigma_{+}) = 0,
\end{eqnarray}
and
\begin{eqnarray}
\partial_{-}f(\sigma_{-}) -  \partial_{+}g(\sigma_{+}) = 0.
\end{eqnarray}
where $\sigma_{\pm} =  \sigma \pm \tau $ and $\partial_{\pm} = \frac{ \partial_\sigma\pm \partial_\tau }{2}$. It implies $f(\sigma_{-}) = c_1$ and $g(\sigma_+) = c_2$, where $c_1$ and $c_2$ are constants. Note that this is not a requirement neither from the equation of motion, nor from the constraints. So demanding a cusp implies a fine-tuning on the initial conditions. In the words of \cite{Spergel:1986uu}, formation of cusps in this scenario would demand a `conspiracy'.

On the other hand, the situation for a superconducting cosmic string can be different. 
The presence of an electromagnetic current would add an extra term to the constraints (\ref{const1}) and (\ref{const2}) making cusp formation more natural in this set up or act as a source term in the equations of motion. To explicitly show that in this set up, one has to generalize the wiggly Polyakov action by including interactions to the electromagnetic field. We leave this for a future work. The constraints coming from FRBs data on superconducting strings parameters, was studied in \cite{Ye:2017lqn}.





\section{Discussion and Conclusion}

We have computed three consistency conditions, to wit, the distance relation, time scale, and emission rate that a cosmic string should pass in order to be a solid candidate as a source of FRBs. We have done so by taking into account the effect that Lorentz boosts have on the cusp length and assuming a weaker assumption for the scaling of the energy. It was shown that even though it is possible to tune the scaling of the energy decay of particles to make the distance relation and time scale compatible with observations, the emission rate is too high in that case. 

A new derivation for the cusp formation condition 
using the wiggly version of the Polyakov action was also provided. It is clear from this condition that the cusp formation is very unlikely in a string with a finite thickness, demanding a fine tune of the initial conditions. We also pointed out that, for a superconducting string, the presence of a charge would weaken the constraint condition for cusp formation, making it more likely in this set up.




\section*{Acknowledgements}

We thank Julien Larena and Bryan Gaensler for valuable discussions, as well as Bryce Cyr for detailed explanation of his work and careful reading of this manuscript and Robert Brandenberger and Guilherme Franzmann for their comments. RC, JEBG and AW acknowledge financial support from the South African Research
Chairs Initiative of the NRF and the DST. JEBG is funded by the National Astrophysics $\&$ Space Sciences Programme (NASSP). Any opinion, finding and conclusion or recommendation
expressed in this material is that of the authors and the NRF does not accept
any liability in this regard.

\newpage 

\bibliographystyle{unsrt}
\bibliography{References}

\end{document}